\documentclass{optica-article}

\articletype{Research Article}

\usepackage{lineno}
\usepackage{multirow}
\usepackage{fancyhdr}
\begin{document}

\title{3D scanning microscopy through scattering surfaces using the optical memory effect}

\author{Harish Sasikumar,\authormark{1 *} Gerwin Osnabrugge,\authormark{1} Kirsten Gerritsma,\authormark{1} Bahareh Mastiani,\authormark{1} and Ivo M. Vellekoop\authormark{1}}

\address{\authormark{1}Biomedical Photonic Imaging Group, Faculty of Science and Technology, University of Twente, P.O. Box 217, 7500AE Enschede, The Netherlands\\
}

\email{\authormark{*}harishs.optics@gmail.com} %% email address of the corresponding author

% use {asbstract*} to suppress the copyright line. Copyright information will be added in production
%\renewcommand{\contentsname}{Outline}
%\tableofcontents
%\newpage

\begin{abstract*}
Wavefront shaping allows light to be focused through scattering objects. However, the wavefront correction found is only valid in a small region called the isoplanatic patch. Here we present a simple approach to extend this isoplanatic patch by shifting and scaling the corrected wavefront appropriately, demonstrating an 8.6-fold increase in the lateral scanning range and a 1.5-fold increase in the axial range through a scattering layer without the need to perform additional wavefront shaping measurements. Our findings agree well with a simple geometrical model that also allows us to extract the effective position of the scattering layer from the measurements.
\end{abstract*}

%%%%%%%%%%%%%%%%%%%%% MAIN TEXT %%%%%%%%%%%%%%%%%%
%%%%%%%%%%%%%%
\section{Introduction}

In the field of biomedical imaging, the scattering of light presents a major challenge that limits the achievable resolution and imaging depth. 
Wavefront shaping (WFS) \cite{vellekoop_focusing_2007, kubby_wavefront_2019} and adaptive optics \cite{ kubby_adaptive_2013, ji_adaptive_2017, zhang_adaptive_2023} have emerged as powerful techniques to mitigate the effects of scattering. By using a spatial light modulator to shape the incident wavefront, the distortions introduced by scattering can be effectively counteracted, leading to improved resolution and increased image quality.  

There are many ways to find a correction that optimizes the focus at a given point using deformable mirrors or spatial light modulators (SLMs)\cite{ kubby_wavefront_2019, kubby_adaptive_2013}. Most of these methods are relatively time-consuming and deplete some of the photon budget available for imaging. Moreover, the correction that is found only optimally compensates for distortions along a designated path. Fortunately, there are instances where the same correction can be repurposed for a range of focal points within an imaging volume due to the phenomenon of lateral and axial memory effects \cite{freund_memory_1988, feng_correlations_1988, ghielmetti_scattered_2012, osnabrugge_generalized_2017}.  This volume is called the isoplanatic patch, or isoplanatic volume.

To make optimal use of the optical memory effect, the SLM should be conjugated to a plane at a depth of 1/3 of the propagated distance in scattering samples\cite{osnabrugge_generalized_2017, zhu_chromato-axial_2020}. For samples where scattering occurs in a single plane, e.g. at the sample surface, the isoplanatism is optimized when the SLM is conjugated to this scattering layer \cite{vellekoop_scattered_2010, mertz_field_2015, park_high-resolution_2015, li_conjugate_2015, may_simultaneous_2021}. By directly projecting the corrections onto the primary scattering layer, the correction pattern remains fixed relative to the scattering structure during lateral scanning of the focus.  Arrangements with added remote focusing lenses can be used to maintain the conjugation between the correction pattern and the scattering layer during three-dimensional scanning \cite{tao_three-dimensional_2017}. Recently, other arrangements have utilized an additional scan mirror to stabilize the correction pattern relative to the scattering structure \cite{papadopoulos_dynamic_2020}.

Unfortunately, the optical design for a sample-conjugate arrangement can be quite challenging, especially since the conjugation plane needs to be adjusted on a per-sample basis. Pupil-conjugated arrangement for wavefront shaping is simpler to implement because it requires fewer components and has a simpler optical design \cite{kubby_adaptive_2013}. However, it has a smaller isoplanatic region, meaning that a single correction works only "near" the original focus. 

In this work, we explore a simple scheme to maximize the isoplanatic volume in a pupil-conjugate configuration. We have developed a geometrical model predicting that by scaling and shifting a calculated wavefront correction, it can be repurposed for regions different from the initially calculated volume. Additionally, we have formulated analytical models to predict the expected degradation in image enhancement due to such reuse of correction patterns.

We validated our method by imaging a sample with a scattering surface in a two-photon microscope with a pupil-conjugate SLM. We demonstrated a 8.6-fold increase in lateral scanning range and a 1.5-fold increase in the axial scanning range by shifting and scaling the correction pattern on the SLM. These increases in scan range amount to a more than a 100-fold increase in the isoplanatic volume.

\section{Principle}

\begin{figure}[htb]
\centering
\includegraphics[width=\textwidth]{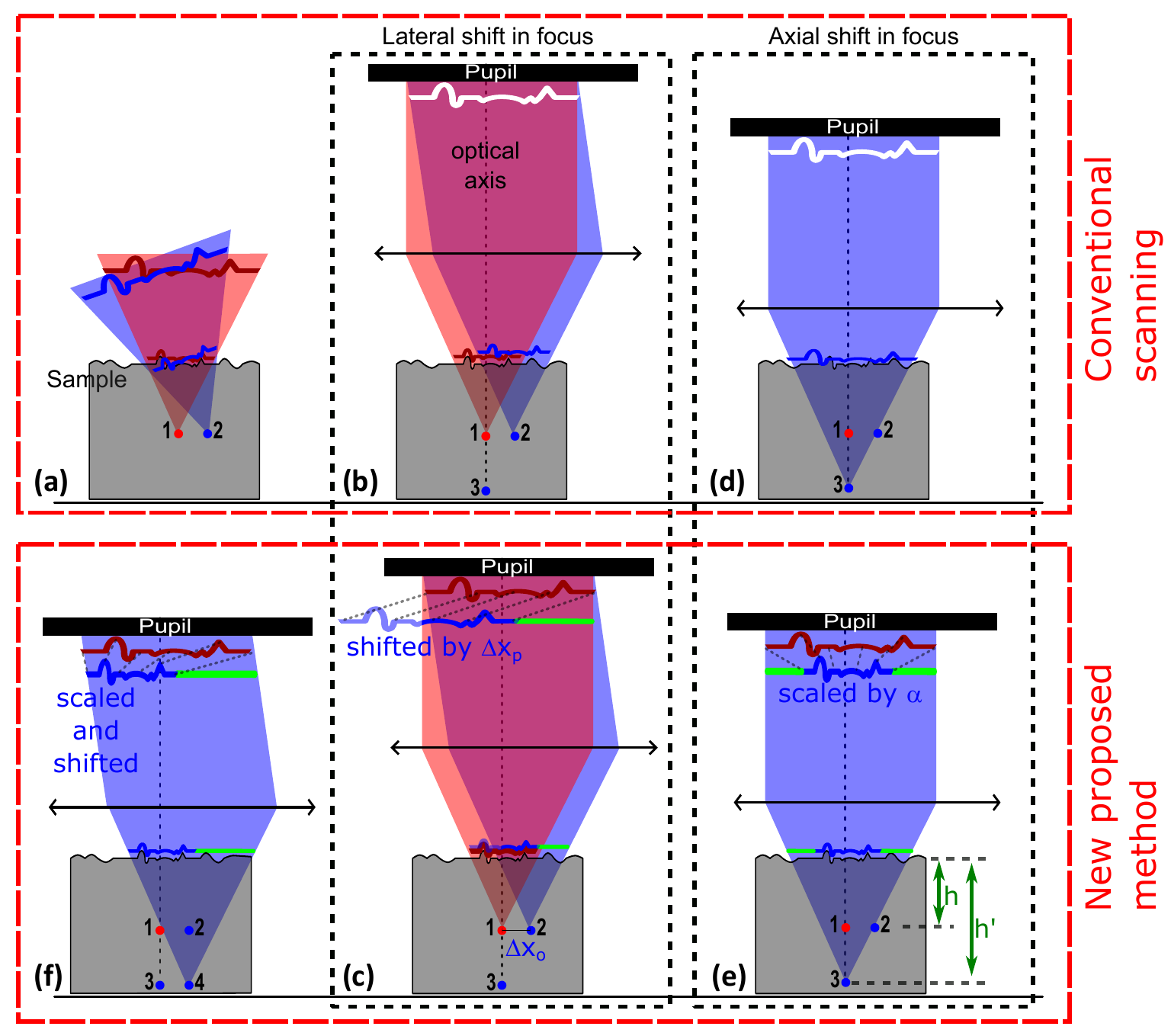}
\caption{ 
(a) Schematic depicting the optical memory effect in a sample with a scattering top surface. For simplicity in the schematics, we assume the refractive index to be uniform everywhere, except at the scattering surface. As the beams pass through identical regions in the scattering layer, the correction pattern can be reused. (b) Lateral scanning of the focus, resulting in a corresponding shift in the wavefront correction. 
(c) Wavefront corrections are shifted in the pupil plane to counteract their shift during lateral scanning. 
(d) Axial scanning of the focus, by moving the objective closer to the sample.
(e) Scaling of the correction pattern in the pupil plane to match with the axially shifted scattering layer. 
(f)  A combination of scaling and shifting in the pupil plane when the focus undergoes both axial and lateral displacements.}
\label{im:method}
\end{figure}

The optical memory effect is a phenomenon that enables the reuse of wavefront corrections applied to focus light at one point for focusing at another point. In the context of imaging through a scattering layer, Fig.~\ref{im:method} (a) illustrates this effect. Here, a wavefront correction designed for point 1 can be applied to focus on point 2 as both the beams propagate through same region of the scattering layer. 

However, this method cannot be directly used with a scanning microscope in a pupil-conjugate arrangement. As the focus is scanned laterally, the wavefront corrections also move with respect to the sample, as illustrated in the laterally shifted beam in Fig.~\ref{im:method} (b).  The laterally shifted correction pattern no longer corrects for the scattering encountered by the beam. Our solution to this problem is to shift the correction pattern displayed on the SLM, resulting in a shift of the pattern in the pupil plane as illustrated in Fig.~\ref{im:method}(c). As derived in the supplementary material (section 1), the correction at the pupil needs to to be displaced by 
\begin{equation}
    \Delta x_{p} = \Delta x_o\frac{n f_o}{n_w h}
    \label{eq:x_slm}
\end{equation}
with $\Delta x_o$ being the lateral shift of the focus during scanning, $n$   the refractive index of the sample, $f_o$  the focal length of the objective, and $h$ the distance between the imaging plane and the scattering layer.

With this shift, the correction pattern remains laterally aligned with the scattering surface during scanning. Note that there are  regions in the shifted correction pattern that are now located outside of the pupil-stop and therefore do not contribute to the wavefront shaping. Conversely, there are regions in the pattern that cannot be solely derived from the initial correction pattern. These portions are indicated by uniform, flat regions in the correction wavefront.

A similar situation occurs when scanning the focus axially. As illustrated in Fig.\ref{im:method}(d), if we increase the imaging depth by moving the objective closer to the sample, the correction pattern scales relatively to the scattering surface. We assume that the pattern is sufficiently smooth to neglect diffraction effects so that we can compensate for this change in distance by simply scaling the pattern on the SLM (Fig.\ref{im:method}(e)). As derived in the supplementary material (section~1), the scaling needs to be about the optical axis with a factor,

\begin{equation} \alpha = \frac{h}{h'} \label{eq:alpha} \end{equation}

where $h'$ is the depth below the scattering layer we want to image, using a correction that was realized to focus at a depth $h$. 

In cases involving both axial and lateral displacement of the focus, the pattern needs to undergo a combination of scaling and shifting at the pupil of the objective, as depicted in Fig.~\ref{im:method}(f).

\section{Setup and data processing}

\begin{figure}[ht]
\centering
\includegraphics[width=\textwidth]{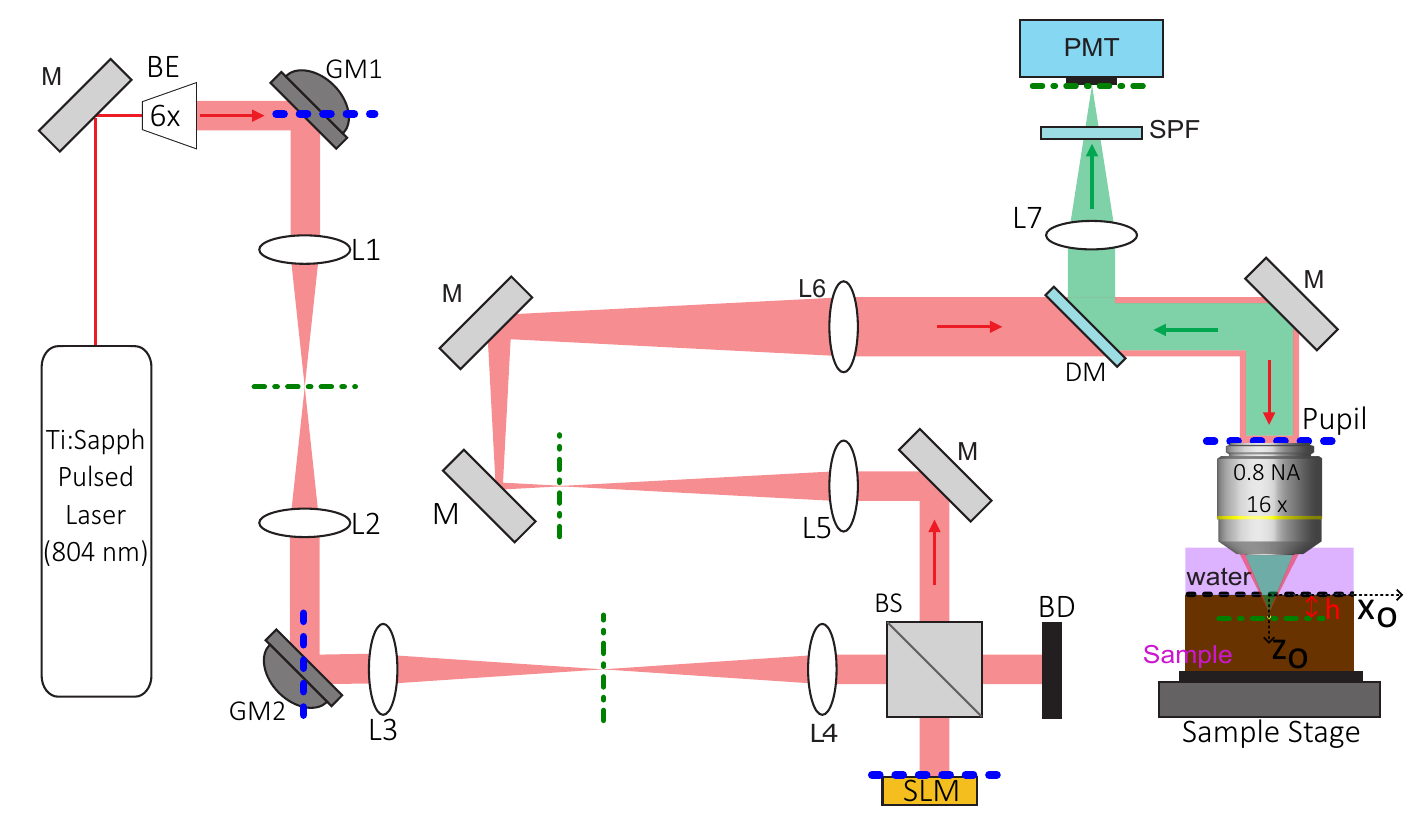}
\caption{ 
Schematic of our two-photon excitation microscope setup, showing key components: laser source, beam expander, galvo mirrors (GM1,2), spatial light modulator (SLM), photomultiplier tube (PMT) and mirrors (M). Lenses (L1-7) are used for creating conjugate planes of the image (broken green lines) and pupil (broken blue lines). Lenses L5 ($f_5=150$ mm) and L6 ($f_6=300$ mm) are used to conjugate the SLM to the pupil plane of the objective. Dichroic mirror (DM) and a short pass filter (SPF) are used for separating fluorescence signal from the excitation. Details such as additional folding mirrors, power and polarization controllers, and periscopes are omitted for clarity.}
\label{im:setup}
\end{figure}

Figure~\ref{im:setup} presents a schematic of our two-photon excitation microscope setup. The excitation is accomplished using a titanium-sapphire laser (Spectra-Physics, Mai Tai) operating at a wavelength of 804 nm. The laser beam is expanded $6\times$ and subsequently directed towards two galvo mirrors (GM1 and GM2) with orthogonal axes of rotation, and finally to a calibrated \cite{cox2025inline} phase-only spatial light modulator (SLM, LC, Meadowlark Optics, 1920 × 1152 pixels). Relay lenses conjugate the galvo mirrors and the SLM to the pupil plane of the water immersion objective lens (Nikon, CFI75 LWD 16x, NA 0.8, with a focal length $f_o=12.5$ mm). Note that the focal lengths of $f_5$ (150 mm) and $f_6$ (300 mm) are chosen such that the SLM pattern is demagnified by a factor of 2x in the objective pupil plane. A dichroic mirror (DM, Semrock, FF685-Di02-25×36) and a short pass transmission filter (SPF, Semrock, FF01-680/SP-25) separate the emission from the excitation light. The emitted light is finally detected by a photomultiplier tube (PMT, Hamamatsu, H10770(P)A-40/-50).

Scanning of the focus along the lateral direction through the sample is done by deflecting the beams using the galvo mirrors (GM1 and GM2). Lateral scanning of the SLM pattern shifts the wavefront correction with respect to the beam center, but does not move the focus position. Scanning of the focus along the axial direction (along $z_o$ axis) is done by moving the objective using a piezo scanning stage (PI, PD72Z2x/4x). 

We utilized a polydimethylsiloxane (PDMS) sample (refractive index, $n = 1.41 \pm 0.005 $) containing fluorescent beads (Fluoresbrite, plain YG) of diameter 500 nm. The sample had a rough surface with a 120 grit texture on its top as described in \cite{thendiyammal_model-based_2020}. This top layer acted as a diffusive layer so that the beads were clearly visible only up to a depth of $<$100  µm.

To evaluate the quality of the images obtained during our experiments, we calculated the root mean square of the pixel values, which is a good metric for assessing image quality \cite{sasikumar_metal-enhanced_2020, cakir_contrast_2018}. After removing the background signal, we measured the improvement in image quality by comparing the quality ratios of images with and without wavefront shaping. Further details can be found in the supplement, section~3.

\section{Results and discussions}

\subsection*{Lateral extension of scanning range}

\begin{figure}[htb!]
\centering
\includegraphics[width=\textwidth]{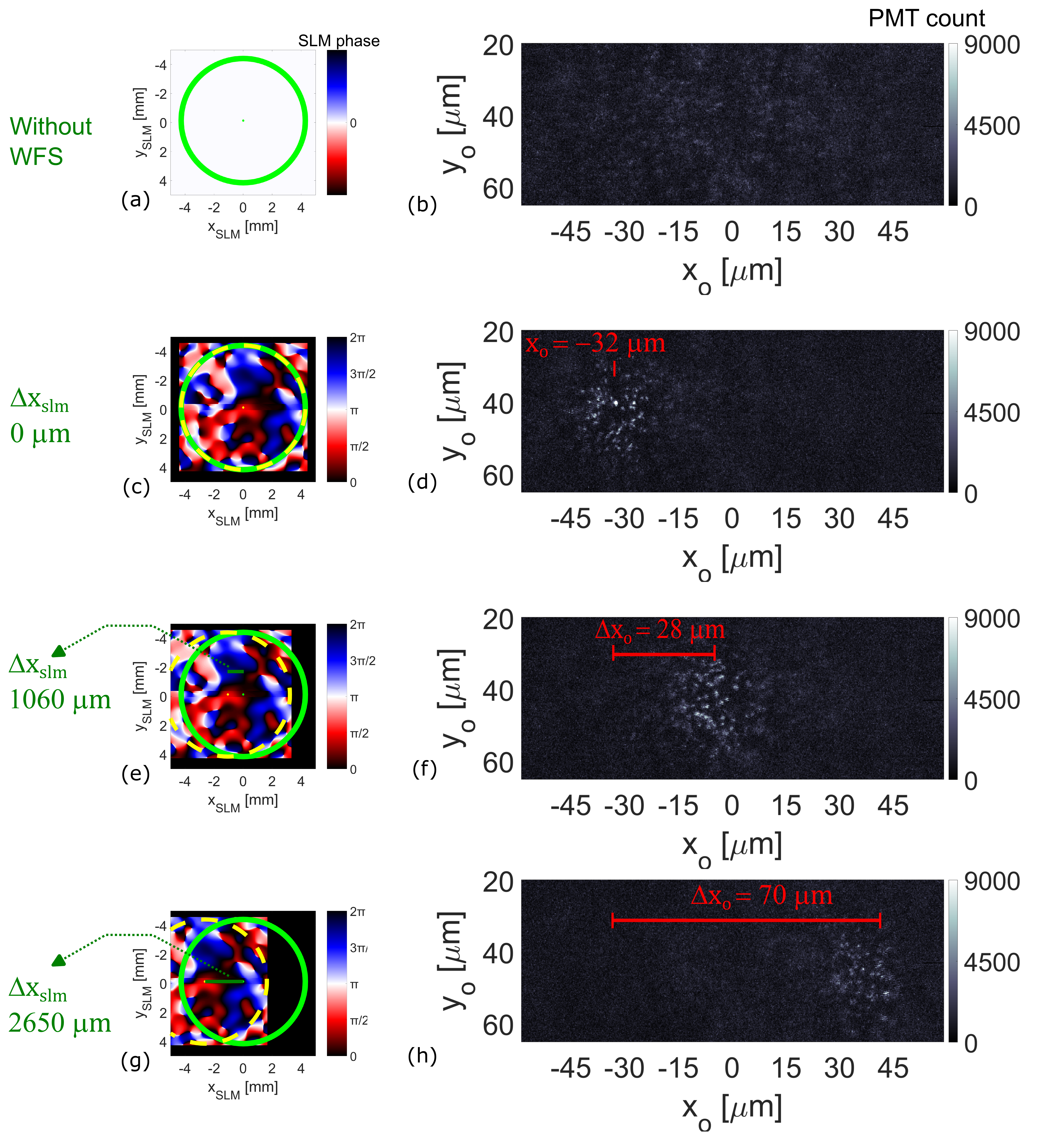}
\caption{Experimental results obtained during wavefront shaping and lateral shift of the shaped patterns. (a) Uniform phase pattern without WFS, with the green circle and dot indicating the mapping of the pupil aperture at the SLM and its center, respectively (b) Corresponding image without any correction. (c) and (d) are respectively, the wavefront correction as the SLM phase pattern and the resulting image. (e) and (g) are the SLMs with the patterns shifted by 1060 µm and 2650 µm, respectively. (f) and (h) represent the corresponding images with the isoplanatic patch displaced by 28 µm  and 70 µm.}
\label{im:lateral_me_images}
\end{figure}

We begin the experiments by imaging the sample. Figure~\ref{im:lateral_me_images}(b) displays measurements without any correction, i.e. with a flat wavefront on the SLM (Fig.~\ref{im:lateral_me_images}(a)). Note that the $x_o$ and $y_o$ axes represent the lateral dimensions, and the $z_o$-axis is the orthogonal direction along the depth of the sample. The beads are barely visible due to low signal intensity caused by scattering. 

Next, we performed wavefront shaping to maximize the intensity at a specific bead (at location $x_o$ = -32 µm, $y_o$ = 40 µm). We used a Fourier-based optimization algorithm \cite{mastiani_wavefront_2022} that is suitable for forward scattering samples. The resulting phase pattern of the correction wavefront was then projected onto our SLM (Fig.~\ref{im:lateral_me_images} (c)), and the corresponding image in the object plane was recorded (Fig.~\ref{im:lateral_me_images} (d)). The intensity of the fluorescent beads is enhanced by a factor of 6.8, and the quality within the isoplanatic patch is enhanced by a factor of 1.9 in the neighbourhood of the bead.

We then shifted the pattern on the SLM and captured corresponding images in the object plane. We observed a 28±2 µm  shift in the isoplanatic patch (Fig.~\ref{im:lateral_me_images}(f)) for a pattern shift of 1060 µm on the SLM (Figure~\ref{im:lateral_me_images}(e)). From Eq.~\ref{eq:x_slm}, we predicted a linear relationship between the shift in the isoplanatic patch and the displacement in the conjugate pupil plane. This was observed in Fig.~\ref{im:lateral_me_images}(e-h), where larger shifts in the SLM showed a proportionately larger shift in the location of the isoplanatic patch. 

\begin{figure}[ht]
\centering
\includegraphics[width=0.8 \textwidth]{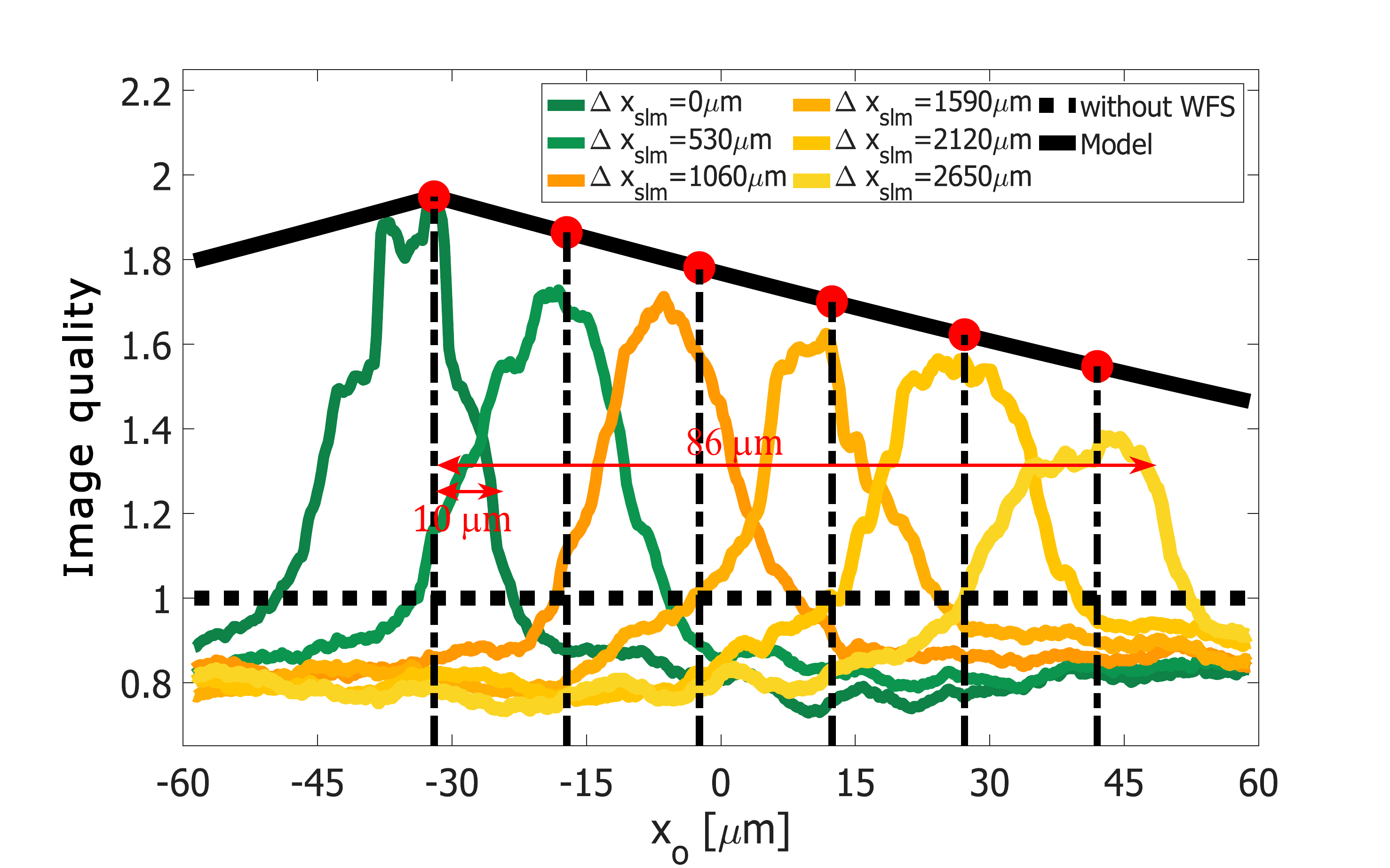}
\caption{Variation in image quality enhancement and scanning range with lateral displacement of the correction pattern at the SLM. In experiments, image quality is computed for volumes with dimensions 7.6 µm $\times$ 25 µm $\times$ 21.2 µm along the $x_o, y_o $ and $z_o$ directions. The theoretical curve is an estimate of the image quality with continuous shifting of patterns (details in the supplementary information, section~4). The vertical lines indicate the expected locations of the regions of interest fitted to our model (Eq.~\ref{eq:x_slm}) for the discrete shifts in patterns used in the experiments.}
\label{im:lateral_memory_effect}
\end{figure}

Equation~\ref{eq:x_slm} can be used to estimate the distance between the scattering layer and the imaging plane, $h$, by linearly fitting the shift in the isoplanatic patch (regions of enhanced quality in Figure~\ref{im:lateral_memory_effect}) and the displacement in the conjugate pupil plane. By fitting theses displacements, we found that the factor $\frac{nf_o}{n_w h}$ in Eq.~\ref{eq:x_slm} to be 71.6 ± 1.5. Taking the refractive indices of of PDMS and water to be $n=$ 1.41 and $n_w=$ 1.33, respectively and the focal length to be $f_o=$ 12.5 mm, we estimated $h$ to be between 185±4 µm.

We independently validated the distance to the scattering layer by imaging the scattering layer. After focusing at the highest point of the scattering layer, the objective was displaced by 190 µm to reach the imaging plane. Considering a measured roughness of approximately 28 µm in the scattering layer, the effective displacement relative to the sample surface ranges from 162 to 190 µm. Using an online database\cite{polyanskiy_refractive}, we obtained the refractive indices of water (1.33)\cite{daimon_measurement_2007} and PDMS (1.41) \cite{zhu_surface-plasmon-resonance-based_2017} at 804 nm. With these values and Eq.~S7 in the supplement the distance between the scattering layer and the image plane was estimated to be between 172 µm and 201 µm, which agrees well with the depth measured using the model.

Figure~\ref{im:lateral_memory_effect} presents the quality enhancement for different lateral displacement of the correction pattern at the SLM ($\Delta x_\textrm{slm}$). A unit quality improvement corresponds to the improvement in quality for an image  obtained without any wavefront shaping. This reference level is denoted by the black horizontal dotted line. The highest quality is obtained in the image when applying wavefront shaping without any shifting at the SLM ($\Delta x_\textrm{slm} = 0$). As expected, this is observed  around the bead at which the optimization was performed. It is noteworthy that, when using an unshifted wavefront, the quality deteriorates below the reference level, as we image farther away from the optimized bead. This is attributed to the significant lateral displacement causing a complete mismatch between the correction pattern and the scattering layer. Consequently, the correction pattern acts as an additional phase "distorter" that even decreases the quality of the focus compared to a flat wavefront. 

There is a limit to how far the pattern on the SLM can be shifted, since the portions of correction pattern shifted outside of the pupil (Figure~\ref{im:lateral_me_images} e, g) is useless in wavefront shaping. This limits the maximum isoplanatic volume that can be achieved by shifting the correction patterns. This also reduces the image quality  as we increasingly shift the pattern outside of the pupil plane. 
We developed a model that predicts the variation in quality improvement as we shift the correction pattern (Eq.~S10 in the supplement). Experimentally, we could observe a similar reduction in the image quality as we progressively shifted the pattern on the SLM (Fig.~\ref{im:lateral_memory_effect}).

Using this prediction we could also estimate the increase in isoplanatic area provided by our method of pattern shifting. Experimentally, if we define the scanning range as the region where the quality improvement reaches a minimum of 1.2, we can see that the isoplanatic patch can be extended from 10 µm to 86 µm. This 8.6-fold increase in the two lateral dimensions amounts to a factor 73 increase in the area of the isoplanatic patch.

\subsection*{Axial extension of scanning range}

From the previous measurements, we know the imaging depth to be around 185 µm. Equation (\ref{eq:alpha}) predicts how to scale the correction in order to image deeper into the sample. We predicted that scalings of 0.9 and 0.8 are optimally suited to image at depths of 205.6 µm and 231.2 µm, respectively. To verify this, we scaled the pattern to 0.9 and 0.8 times its original dimensions and acquired images at different depths.

\begin{figure}[ht]
\centering
\includegraphics[width=\textwidth]{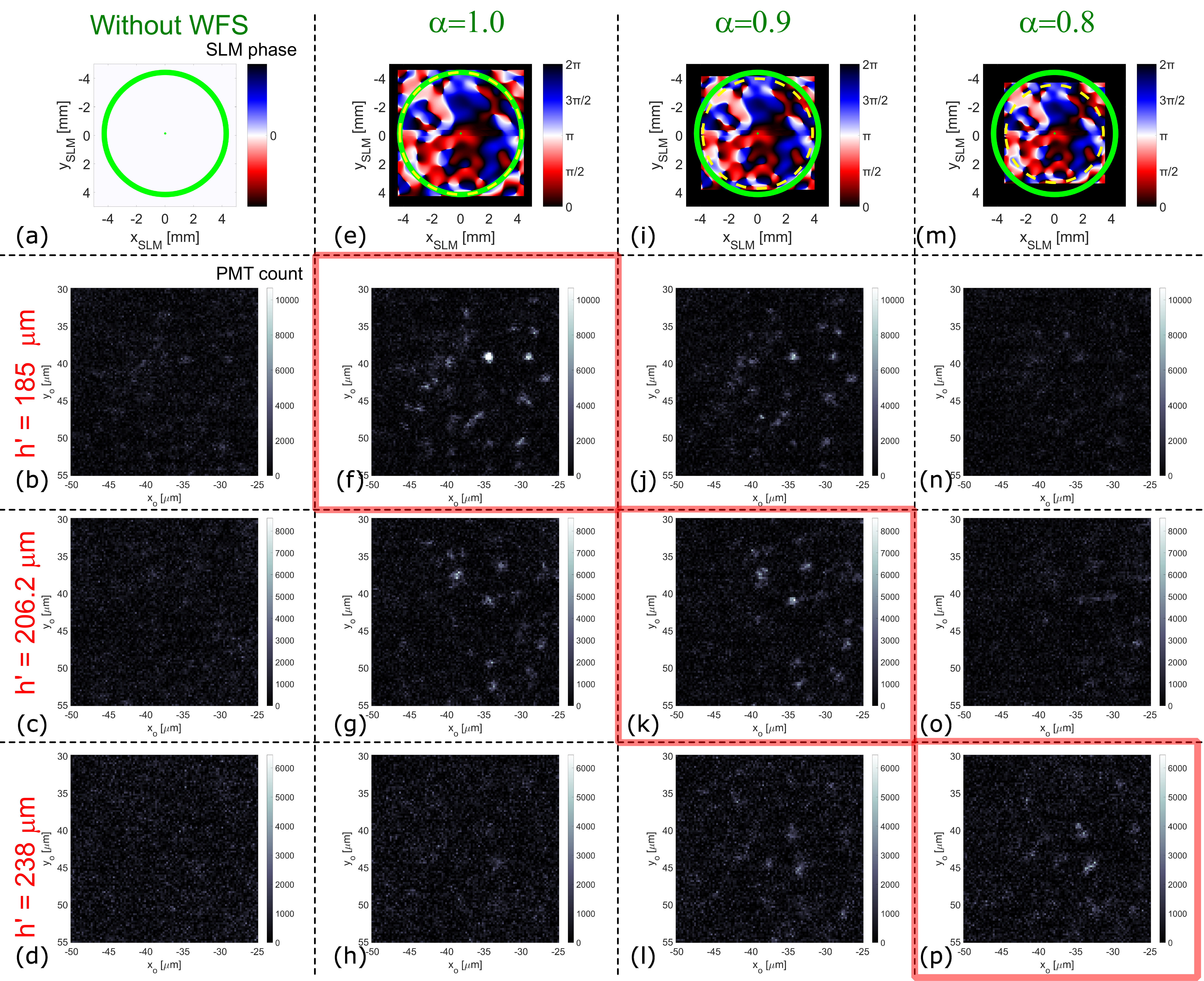}
\caption{Applying different wavefront scaling ($\alpha$) for imaging at varying depths. The SLM patterns are shown as (a) without any wavefront shaping (e) wavefront shaped without any scaling ($\alpha=1$) and optimized for $h=185$ µm (i) with $\alpha=0.9$ (m) with $\alpha=0.8$. The green circle and dot indicates the mapping of the pupil aperture at the SLM and its center, respectively. (b)-(d), (f)-(h), (j)-(l) and (n)-(p) are the images taken with these SLM patterns and at depths of $h'=$ 185 µm, 206.2 µm and 238 µm below the scattering surface. Notably, the highest image quality near the imaging depths of $h'=206.2$ µm and $h'=238$ µm is achieved when utilizing scaling factors of $\alpha=0.9$ and $\alpha=0.8$, respectively, following the trend predicted by Eq.~\ref{eq:alpha}. }
\label{im:axial_me_images}
\end{figure}

Figure~\ref{im:axial_me_images} illustrates typical images obtained during these measurements. Figure~\ref{im:axial_me_images} (a) represents a flat wavefront at the SLM. Figures~\ref{im:axial_me_images} (b), (c), and (d) show the images captured at depths of 185 µm, 206.2 µm, and 238 µm, respectively, using only a flat wavefront at the SLM. 

Then, a correction wavefront was shaped at 185 µm depth, as depicted in Figure~\ref{im:axial_me_images} (e). Corresponding images with the correction wavefront projected at the SLM and at different depths are presented in Fig.~\ref{im:axial_me_images} (f-h). Further, the SLM pattern was scaled by a factor of  $\alpha=0.9$ (Figure~\ref{im:axial_me_images} (i))  and $\alpha=0.8$ (Figure~\ref{im:axial_me_images} (m)) and images were taken again at varying depths. We can see that higher intensity images are  obtained for deeper depths with smaller scaling.

\begin{figure}[ht]
\centering
\includegraphics[width=0.8\textwidth]{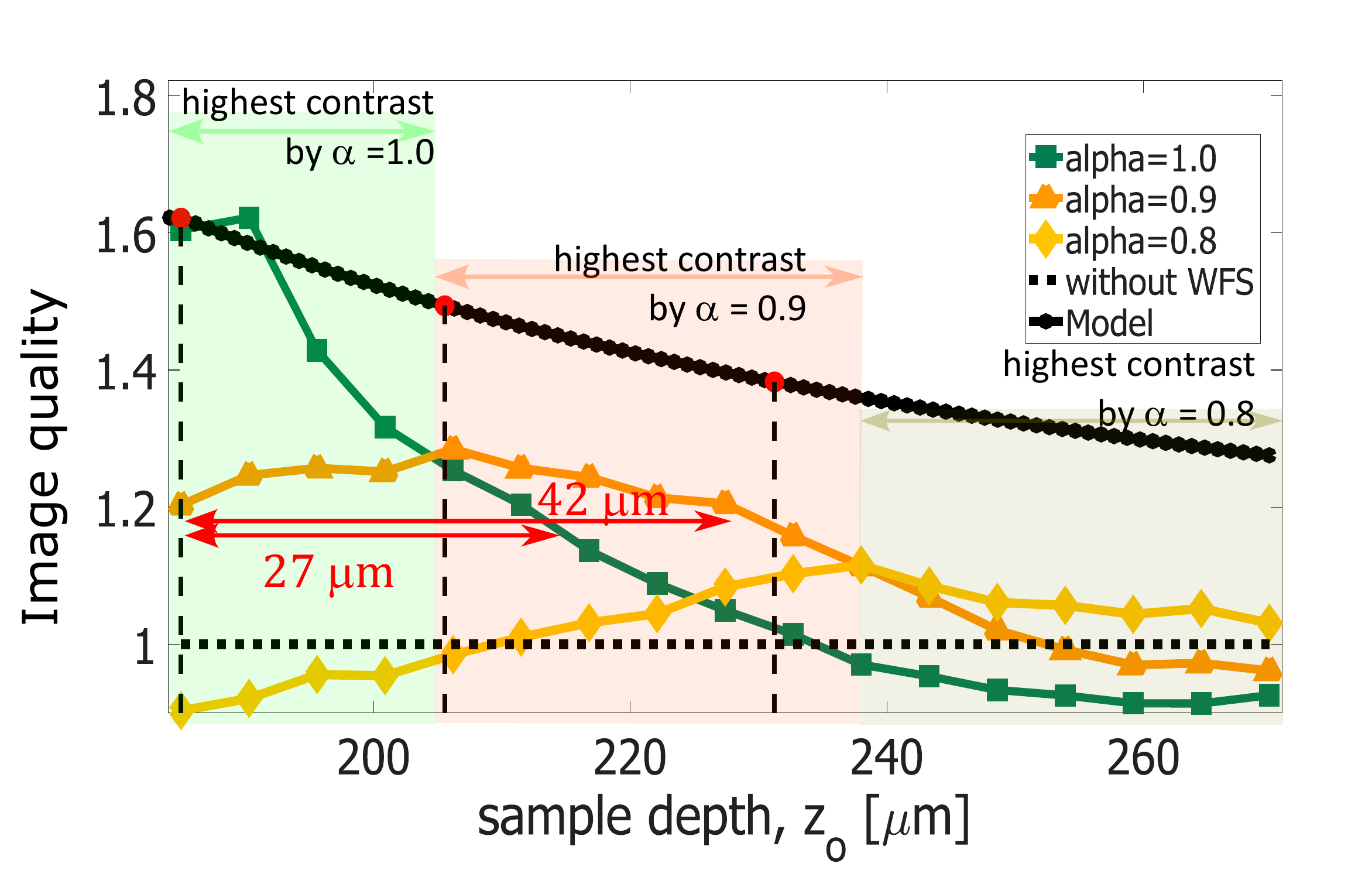}
\caption{Impact of wavefront scaling ($\alpha$) on image quality at varying depths below the scattering surface. Quality is computed for larger volumes with dimensions 25 µm $\times$ 25 µm $\times$ 16 µm along the $x_o, y_o $ and $z_o$ directions. The maximum image quality, when $\alpha$ = 1, is obtained up to 204 µm, determined by the cross-over point of the experimental curves. For $\alpha = 0.9$, the range is from 204 µm to 238 µm and $\alpha = 0.8$ is most suited for depths larger than 238 µm.}
\label{im:axial_memory_effect}
\end{figure}

Figure~\ref{im:axial_memory_effect} showcases a more comprehensive picture of the influence of wavefront scaling ($\alpha$) on image quality enhancement at various imaging depths beneath the scattering surface. The dotted horizontal line represents the reference level of unit quality improvement, indicating an image without wavefront shaping. The remaining curves correspond to the image quality distribution achieved with wavefront shaping using the same wavefront but with different scaling. Among the scaling used in this study, an unscaled wavefront ($\alpha = 1$) yields the highest image quality only up to a depth of 204 µm. For depths ranging from 204 µm to 238 µm, a scaling factor of $\alpha = 0.9$ provides the optimal image quality enhancement. This closely matches the theoretical prediction according to Eq.~\ref{eq:alpha} which recommends a scaling of 0.9 at a depth of 205.6 µm. Similarly, the model predicts a scaling of $\alpha = 0.8$ for a depth of 231.2 µm. This is close to the 238 µm where we observed that the scaling $\alpha = 0.8$ starts outperforming the other scalings used in the experiments. 

Scaling down the correction pattern reduces the effective area of the correction wavefront that fills up the pupil, leading to a drop in image quality. We developed a simple model (Eq.~S11 in the supplement) based on paraxial ray propagation to estimate this drop. As shown in Fig.\ref{im:axial_memory_effect}, the measured image quality drops faster than the theoretical expectation, indicating that the assumptions in our model are not fully valid. Still, the model can be used to estimate an upper limit for the expected quality.

In conclusion, scaling the correction pattern down increases the image quality at larger depths, as predicted. If we consider a quality improvement of at least 1.2 as the cut-off, we observe that without any scaling in the pattern the axial scanning range  is limited to 27 µm. By introducing pattern scaling, the scanning range expands to 42 µm along the axial direction, an increase by a factor of more than 1.5.

\section{Conclusions}

In conclusion, we have demonstrated a versatile method that allows a single correction wavefront to be effectively utilized across various lateral and axial positions in the object plane. This approach simplifies the imaging process by significantly expanding the imaging volume accessible with a single correction wavefront in a straightforward, pupil-conjugate microscope setup.

Through the application of geometrical optics-based models, we have derived the necessary shift and scaling that is required to extend the range over which a single wavefront correction is valid. Furthermore, we have developed analytical models that describe the observed reduction in imaging quality when the correction wavefront is shifted and scaled. Further, we have experimentally validated our models using a pupil-conjugated microscopy setup. We showed the ability of our approach to extend the lateral scanning range by a factor of 8.6 and the axial range by 1.5 in our scattering sample, resulting in a 100-fold increase in isoplanatic volume with relatively little effort. The correlation between the predictions of these models and our experimental findings confirms the reliability of our approach and provides a simple way to understand the underlying phenomena affecting imaging performance.  

The increase in isoplanatic volume means that far fewer correction wavefronts need to be constructed, saving a proportionate amount of time and photon budget. Ideally, a scanning microscope would scan one isoplanatic volume at a time, with the SLM updating patterns between volumes.  We anticipate that our method can be extended to imaging in volumetric scattering tissues, which can be approximated as a scattering layer at a depth of $L/3$ in perfectly homogeneous scattering samples \cite{osnabrugge_generalized_2017}.  In a practical scenario, the depth of this effective scattering layer could be estimated by shifting the correction pattern using the method described in this work. Finally, the fact that a pupil-conjugate setup can be used reduces the complexity of the setup and greatly increases the compatibility with existing microscope systems.

%%%%%%%%%%%%%%%%%%%%% BACK MATTER %%%%%%%%%%%%%%%%%%%%%%%%%%%%%%

\begin{backmatter}
\bmsection{Funding}
This publication is part of the NWO-I Neurophotonics project with number 16NEPH01 which is financed by the Dutch Research Council (NWO).

\bmsection{Acknowledgments}
The authors gratefully acknowledge Tom Knop for his support with our two-photon setup and Merle van Gorsel and Abhilash Thendiyammal for their assistance with the calibration procedures. HS would like to express gratitude to Michele Gintoli and Daniël Cox for their discussions on optics and programming, and to the BMPI group at the University of Twente for the conversations and valuable inputs on the work.

\bmsection{Disclosures}
\noindent The authors declare no conflicts of interest.

\bmsection{Data Availability}
Measurement data and analysis code are available from 4TU Research Data\cite{4tu_data_sasikumar2026me}. The experiments in this article can be done using our updated Python library \cite{vellekoop2023OpenwfsGithub, doornbos2025OpenWFSLibrary}.

\bmsection{Supplemental document}
See Supplement 1 for supporting content. 

\end{backmatter}

\bibliography{references.bib}

\end{document}

\maketitle

\section{Paraxial ray-optics model}
\label{sec:ray_vector_approx}

\begin{figure}[htbp]
\centering\includegraphics[width=0.58\textwidth]{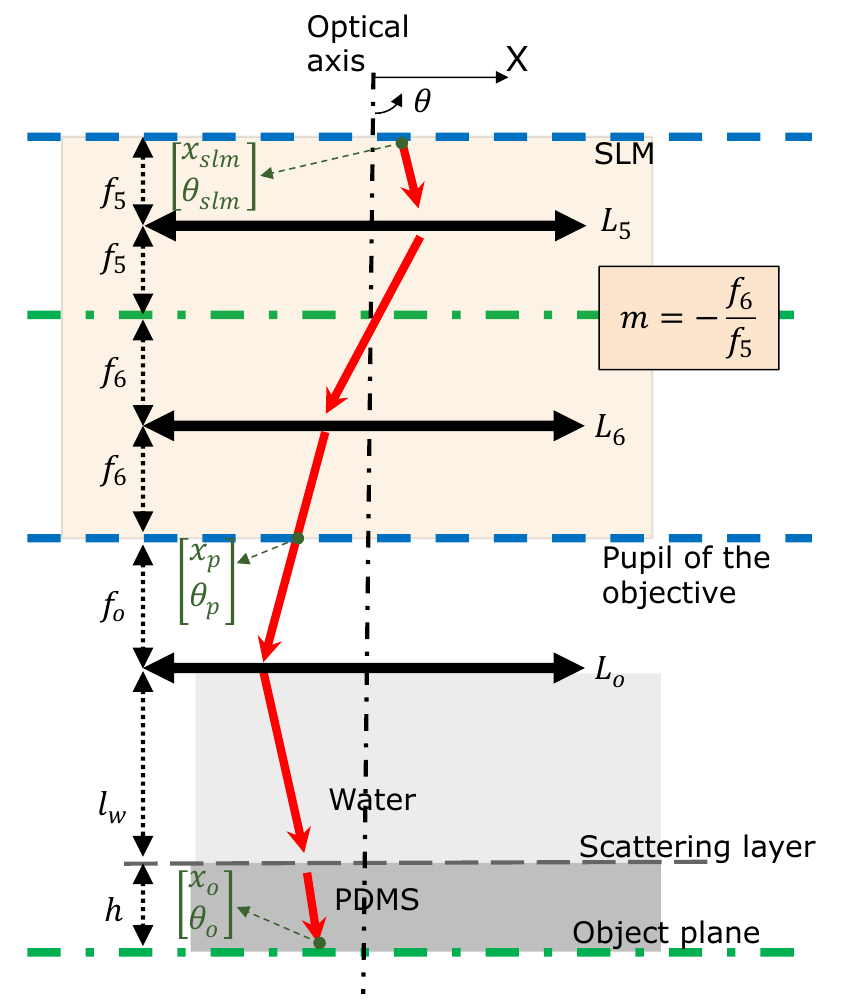}
\caption{ Paraxially approximated light propagation from SLM to the imaging plane. Note that the distances are not to scale. The dashed blue line at the top represents the SLM. The lenses are indicated as $L_x$ with their corresponding focal lengths denoted as $f_x$. Lenses $L_5$ ($f_5=150$ mm) and $L_6$ ($f_6=300$ mm) are used to conjugate the SLM to the pupil plane of the objective with a 4f arrangement of magnification $m$. The objective, $L_o$, has a focal length of $f_o=12.5$ mm. $l_w$ is the height of water below the objective, and $h$ (around 200 µm) is the distance between the imaging plane and the scattering layer
}
\label{SIFig:ray_optics_diag}
\end{figure}

We used paraxial ray optics \cite{hecht_optics_2017} to compute angles and displacements of the rays at different planes of interest in our scanning microscope.  As in the experiment, the spatial light modulator (SLM), functioning as the wavefront shaping element, is conjugated to the back-pupil plane of the microscope objective ($L_o$) through a 4f optical system. The schematic of the portion from the SLM to the focus inside the sample is illustrated in Fig.~\ref{SIFig:ray_optics_diag}. 

In our derivations, we describe rays with column vectors of the form $(x, \theta)^T$, where $x$ denotes the displacement from the optical axis and $\theta$ refers to the counterclockwise angle with respect to the optical axis. The focal length of a lens $L_x$, is represented by $f_x$. We idealized the objective as a thin lens. We modeled the region from the water-immersion objective objective to the focal plane as a sequence of two layers, namely water, and the sample, with refractive indices $n_w$ and $n$, respectively.

\subsection{From SLM to pupil plane of the objective (4-f arrangement)}

The ray vector that emerges from the SLM can be expressed as $(x_\textrm{slm}, \theta_\textrm{slm})^T$. As it propagates through the 4-f arrangement, it is transformed and emerges as a ray vector in the pupil plane of the objective, expressed as $(x_{p}, \theta_{p})^T$. As in a typical 4f arrangement, the displacements at the SLM and the pupil-plane are related by

\begin{equation} \label{eqn:xp}
\begin{bmatrix}
x_p \\ \theta_p 
\end{bmatrix}
= 
\begin{bmatrix}
-(f_6/f_5) x_\textrm{slm} 
\\   -(f_5/f_6)\theta_\textrm{slm}  
\end{bmatrix} 
\end{equation}

Note that the factor $-\frac{f_6}{f_5}$ (= $m$) is the  magnification  between the pupil-plane and its conjugate, provided by the 4f system comprising of the lenses $f_5$ and $f_6$. Thus, any in-plane lateral displacement of the pattern at the SLM results in an equivalent displacement at the pupil with a magnification. Specifically, with our arrangement with $f_6$ = 300 mm and $f_5$ = 150 mm, m= -2. 

The focal length of an infinity-corrected objective is given as \cite{microscopyu_infinity_optics}
$f_o = t_l/M$, where $t_l$ is the tube length and $M$ is the specified magnification. As we used an infinity-corrected, water immersion objective: Nikon, CFI75 LWD 16x, NA 0.8, with a tube length of $t_l=$200 mm and magnification $M=$ 16, the focal length is $f_o=$ 12.5 mm.

\subsection{From the pupil plane to the object plane}

After passing relay lens $L_6$, a light ray travels through the following media before reaching the object plane (see Fig.~\ref{SIFig:ray_optics_diag}:

\begin{itemize}
        \item air (refractive index 1) by a distance $f_o$ to reach the objective lens $L_o$
        \item the microscope objective $L_o$, modeled as a thin convex lens with focal length $f_o$ 
        \item water (refractive index $n_w$) by a distance of $l_w$ to reach the scattering layer
        \item scattering layer as a water-PDMS interface
        \item PDMS (refractive index $n$) by a distance $h$ to reach the object plane
\end{itemize}

Using standard ray-transfer matrices\cite{hecht_optics_2017}, we find the position and angle of the ray in the object plane 

\begin{equation} \label{eqn:x11}
\begin{bmatrix}
x_{o} \\ \theta_{o} 
\end{bmatrix}
= 
\begin{bmatrix}
f_6/(f_5f_on)(h n_w - f_on + l_wn) x_\textrm{slm} - (f_5 f_o/f_6)\theta_\textrm{slm}
\\ f_6 n_w/(n_pf_5f_o) x_\textrm{slm} 
\end{bmatrix} 
\end{equation}

For perfect focusing in the object plane, all rays must converge to the same point, meaning that $x_o$ must be independent of $x_\textrm{slm}$, i.e. $(h n_w - f_on + l_wn)=0$. Thus, the distances and focal length must satisfy the relation

\begin{equation}
\label{eqn:condition_for_focusing}
    \frac{h n_w}{n} + l_w = f_o
\end{equation}

\begin{figure}[htbp]
\centering\includegraphics[width=0.95\textwidth]{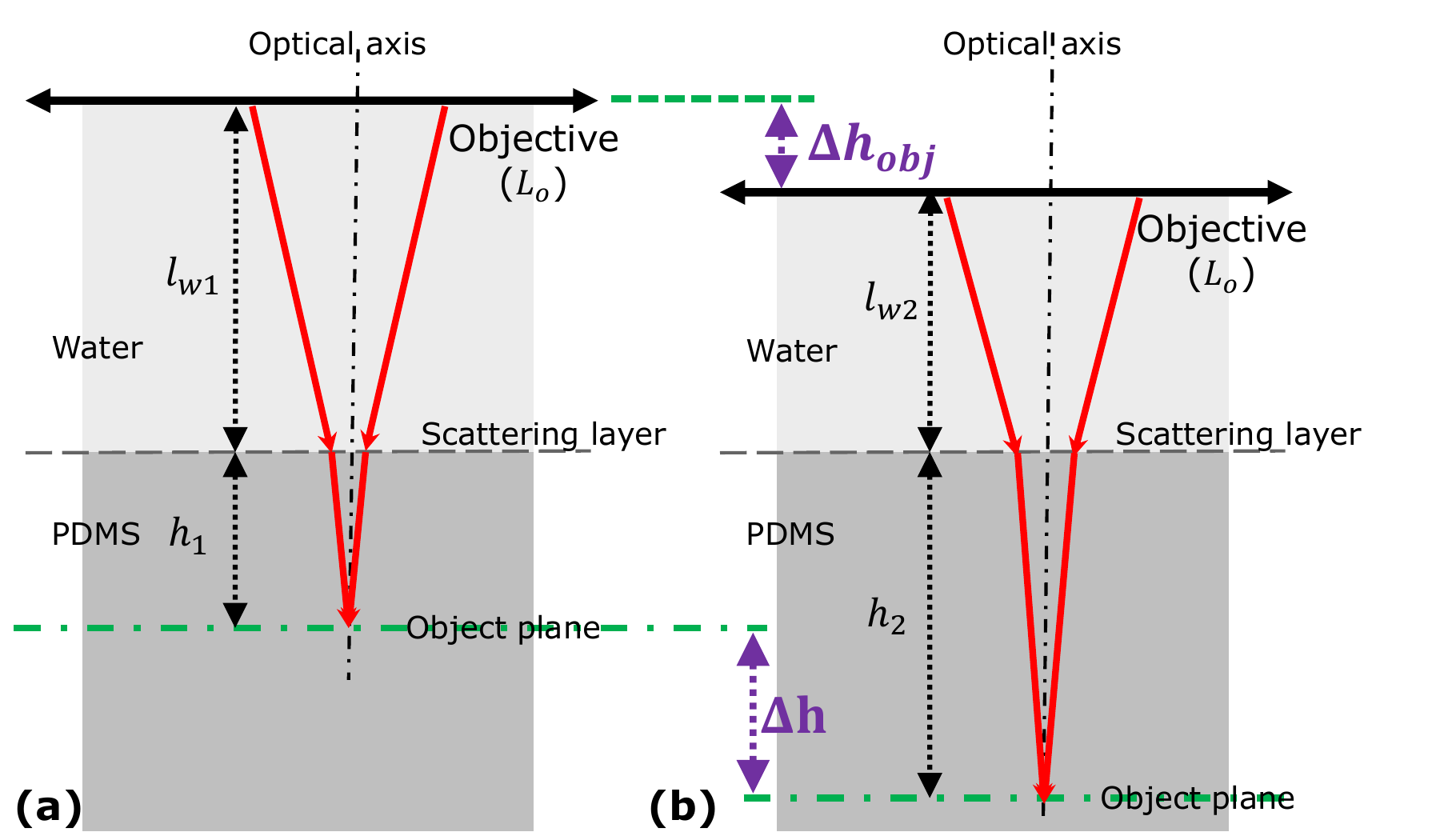}
\caption{ Schematic for measuring the depth of the focus below the scattering layer using the axial displacement of the water-immersion objective.}
\label{SIFig:sample_depth}
\end{figure}

Figure \ref{SIFig:sample_depth} illustrates the implications of this equation, showing that when the objective is brought closer to the sample (and the region of water is replaced with a sample of higher refractive index than water), the focal length moves deeper into the sample through a distance longer than the displacement of the objective.

When the the condition for focusing \ref{eqn:condition_for_focusing}) is met, we can find the position of a ray originating at $x_\textrm{slm}$ in the SLM-plane when it reaches the the scattering layer. This position, $x_{s}$, is given by

\begin{equation}
    x_s = -\frac{f_6 h n_w}{f_5 f_o n} x_\textrm{slm} + x_o
\end{equation}

Thus, the displacements in the SLM ($\Delta x_\textrm{slm}$) and the object plane ($\Delta x_{o}$) are related by

\begin{equation}
   \Delta x_\textrm{slm} = \Delta x_{o}\frac{f_o n}{mhn_w}
    \label{eq:delta_x_slm}
\end{equation}

It is important to note that the axes between the pupil and the SLM are flipped due to the negative sign of the magnification. Additionally, by taking advantage of the circular symmetry around the optical axis, we can extend this relationship to two-dimensional displacements in the pupil and the object planes. In this case, the negative magnification results in an in-plane reflection about the principal point, which is the point where the optical axis intersects with the corresponding lateral plane.

From equation \ref{eq:delta_x_slm}, we can see that the size of the pattern on the SLM and its image on the scattering layer are related through  $h$, the distance from the scattering layer to the focus. To reuse a correction at a different depth $h'$, we should thus scale the SLM pattern by a factor  

\begin{equation}
    \alpha = \frac{h}{h'}
    \label{eqSI:alpha}
\end{equation}

so that the size of image of this pattern on the scattering layer remains constant.

\section{Axial scanning and imaging depth below the scattering layer}
\label{sec:axial_scanning_and_imaging_depth}

In the geometrical optics approximation, the axial displacement of the objective and the corresponding displacement of focus inside a different medium are proportional to the refractive index contrast at the interface \cite{feynman_feynman_2015}. We are deriving this relation for our water-immersion objective
to estimate the depth of the focus below the scattering layer. 

For perfect focusing, Eq.~\ref{eqn:condition_for_focusing} must be satisfied. Moving the objective axially over a distance $\Delta h_\textrm{obj}$  causes $l_w$, the distance propagated through water, to change by the same amount. From  Eq.~\ref{eqn:condition_for_focusing} we now find directly that corresponding displacement of the focus inside the PDMS sample of refractive index $n$ equals:
\begin{equation}
\label{eqn:sample_depth}
     \Delta h = \Delta h_\textrm{obj} \frac{n}{n_w}
\end{equation}

We used this equation to independently determine the imaging depth below the scattering layer, in order to compare with the values obtained by comparing the displacements in Eq.~\ref{eq:delta_x_slm}.  First, we used the microscope to locate the scattering layer itself.  By introducing fluorophores into the water and performing axial scanning of the objective, we observe the fluorescence signal transitioning from a uniform profile above the scattering layer to isolated fluorescent beads deeper into the sample. The scattering layer is identified as the interface between the water and sample where the fluorescence signal abruptly decreases. This locating process achieves an accuracy of approximately 28 µm, which is comparable to the undulations present within the scattering layer itself. Any further displacement of the objective, closer towards the sample, displaces the focus into the sample accoding to Eq.~\ref{eqn:sample_depth} . 

\section{Quantifying improvement in image quality}
\label{sec:quantifying_image_quality}
The root mean square contrast, which is the standard deviation of the pixel values within our regions of interest, provides a quantitative measure of image quality \cite{cakir_contrast_2018}. To quantify the effectiveness of wavefront shaping, we define the image quality enhancement $\eta_c$ as the ratio of contrasts observed in images with and without wavefront shaping. To enable a fair comparison, we compensated for the presence of background noise, which we assumed to be additive and uncorrelated to the signal:

\begin{equation}
    \eta_c = \sqrt{\frac{\sigma^2(I_w) - \sigma^2(I_b)}{\sigma^2(I_{wo}) - \sigma^2(I_b)}}
    \label{eqn:eta}
\end{equation}

Here, the standard deviations $\sigma$ are computed over a volume $V$, which is typically of the order of $10^{-5}~\textrm{mm}^3$. To evaluate image quality in the axial extension, we use a volume of $25~\mu\textrm{m} \times 25~\mu\textrm{m} \times 16~\mu\textrm{m}$ along the $x$, $y$, and $z$ directions, respectively. This volume was chosen to include a sufficient number of microbeads, given the microbead density in our sample, to provide a reliable estimate of the image quality. The images with and without wavefront shaping are denoted as $I_w$ and $I_{wo}$, respectively, while $I_b$ is the background image captured with the laser turned off.

\section{Estimating the performance of reused wavefronts}
\label{sec:estimating_performance}

In our experiments, we reuse a shaped wavefront for different regions-of-interest by shifting and scaling them appropriately. A reused wavefront will perform worse than a wavefront optimized for the new location. This is because, after the transformations on the shaped wavefront, it may only partially fill the pupil of the objective. Thus, there will be a reduction in the focused intensity and a corresponding reduction in performance metrics (like $\eta_c$ as we defined in Eq.~\ref{eqn:eta}) of the wavefront corrections. 

This reduction can be estimated from the area of the used part of the shaped wavefront and can be estimated by calculating the overlap of the modified pattern with the pupil plane. What we are effectively calculating is the overlap integral which quantifies how well the reused wavefront matches the ideally optimized wavefront \cite{Vellekoop_universal_optimal_trans}. 

\begin{figure}[htbp]
\centering\includegraphics[width=0.45\textwidth]{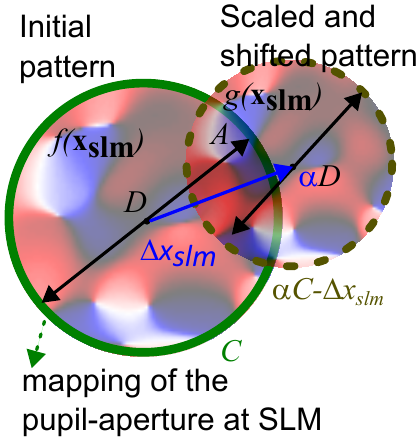}
\caption{ Scaling and displacement of shaped wavefronts in the plane of the SLM }
\label{SIFig:slm_plane}
\end{figure}

As shown in Fig.~\ref{SIFig:slm_plane}, $C$ is a circle representing the pupil aperture mapped onto the SLM plane. It is centered at the origin and has diameter $D$. The original corrected wavefront is denoted by $f(\mathbf{x}_{\mathrm{slm}})$, where $\mathbf{x}_{\mathrm{slm}}=(x_p,y_p)$ is a two-dimensional position vector in the SLM plane. The transformed wavefront is defined as
$g(\mathbf{x}_{\mathrm{slm}})=f(\alpha \mathbf{x}_{\mathrm{slm}}-\Delta\mathbf{x}_{\mathrm{slm}})$,
where $\alpha$ is a scaling factor and $\Delta\mathbf{x}_{\mathrm{slm}}$ is a lateral shift. The transformed wavefront only partially fills $C$ when the transformed circle $\alpha C-\Delta\mathbf{x}_{\mathrm{slm}}$ does not fully contain $C$, i.e., $C \not\subset \alpha C-\Delta\mathbf{x}_{\mathrm{slm}}$. This occurs when $|\alpha|<1$ or when the transformed circle is partially displaced outside $C$.

In the experiments, the objective pupil was overfilled to produce an approximately uniform illumination within the pupil. In addition, only phase corrections were applied, while unfilled pupil regions were assigned a flat phase. We therefore assume that equal pupil areas contribute equally to wavefront shaping and that flat-phase regions do not contribute to quality enhancement.

We assume that the linear intensity response at the focus is proportional to the fraction of the wavefront effectively shaped by the optical system. In non-linear microscopy, the measured quality is related to the signal intensity through a power law with exponent $\chi$. Within this framework, the dependence of the image quality enhancement on the effective overlap area can be modeled as

\begin{equation}
\eta =
\left[
(\eta_0^{1/\chi} - 1)\,A_n + 1
\right]^\chi
\label{eqn:eta_general}
\end{equation}

where $\eta$ is the resulting quality enhancement, $\eta_0$ is the maximum experimentally observed enhancement under full overlap and optimal alignment, and $A_n \in [0,1]$ denotes the normalized overlap area between the mapped pupil aperture at the SLM and the shifted or scaled wavefront. Finally, $\chi$ is the nonlinear exponent relating measured quality to optical intensity in the low-signal regime. In particular, $\chi$ captures the effective power-law response of aberrations on the imaging system, which arises in nonlinear detection processes. Typically, $\chi$ is close to, but slightly less than, 2 for a two-photon microscope~\cite{Sinefeld:15}. In this work, we use $\chi = 1.83$, obtained by fitting the effective nonlinear response of wavefront shaping for the microbeads used in the experiment, as done in \cite{Cox:25}. 

Equation~\ref{eqn:eta_general} is constructed by assuming that the overlap area affects the effective optical intensity enhancement in a linear manner. Since the experimentally measured quality enhancement exhibits a nonlinear dependence on intensity, the enhancement is first mapped to a linearized intensity domain using the exponent $1/\chi$. The overlap factor $A_n$ is then applied in this domain, after which the result is transformed back to the measured domain through the exponent $\chi$. In this way, the model captures both the geometric reduction in wavefront overlap and the nonlinear response of the imaging system.

From Eqn.~\ref{eqn:eta_general} We can see that for full overlap ($A_n = 1$), the maximum enhancement is recovered, $\eta = \eta_0$. For zero overlap ($A_n = 0$), no wavefront shaping contribution remains and the response reduces to the baseline value $\eta = 1$. For partial overlap ($0 < A_n < 1$), the enhancement varies monotonically between these limits, reflecting the gradual loss of effective wavefront correction as the overlap decreases.

\subsection*{Lateral extension of scanning range}

In the absence of wavefront scaling ($\alpha=1$), the overlap facture $A_n$ can be estimated by computing the normalized overlap between the mapped pupil aperture at the SLM and the laterally shifted wavefront. For two identical circular apertures of diameter $D$ separated by a lateral displacement $\Delta x_{\mathrm{slm}}$, the normalized intersection area is given by \cite{weisstein_circle-circle}

$$
A_n =
\frac{2}{\pi}
\left[
\cos^{-1}(d_n)
-
d_n\sqrt{1-d_n^2}
\right],
$$

where 
$$
d_n = \left|\frac{\Delta x_{\mathrm{slm}}}{D}\right|
$$

is the normalized separation between the aperture centers. The expression is valid for the range of  $\Delta x_{\mathrm{slm}=0}$ (complete overlap) to $\Delta x_{\mathrm{slm}} =D$ (no overlap).

Using this geometric estimate and Eq.~\ref{eqn:eta_general}, we model the variation in image quality enhancement as the correction pattern is laterally displaced:

\begin{equation}
\eta =
\left[
(\eta_0^{1/\chi} - 1)\,\frac{2}{\pi}
\left[
\cos^{-1}(d_n)
-
d_n\sqrt{1-d_n^2}
\right] + 1
\right]^\chi
\label{eqn:eta_lateral_me}
\end{equation}

\subsection*{Axial extension of scanning range}

We estimate the image quality enhancement under uniform scaling of the correction pattern. For a non-shifted wavefront ($\Delta \mathbf{x}_{\mathrm{slm}} = 0$), the overlap area scales as the square of the scaling factor, such that $$A_n = \alpha^2$$ for $\alpha < 1$. Using Eq.~\ref{eqn:eta_general}, the corresponding image quality enhancement is modeled as
\begin{equation}
\eta =
\left[
(\eta_0^{1/\chi} - 1)\,\alpha^{2} + 1
\right]^\chi.
\label{SIeqn:eta_axial}
\end{equation}

This expression predicts that the enhancement reaches its maximum value $\eta_0$ at $\alpha = 1$, and remains above unity for $\alpha < 1$, corresponding to partial correction of the effective pupil region.

For $\alpha > 1$, the correction pattern is effectively magnified, corresponding to an oversampling of the measured wavefront. In an ideal system with perfect sampling, such scaling would not degrade performance. However, in practice, the finite spatial resolution of the measured correction pattern introduces reconstruction errors that become increasingly significant under oversampling, leading to a gradual reduction in the achievable enhancement relative to the ideal case.

For comparison with experiments, we use the same approach for $\chi$ and $\eta_0$ as in the analysis of lateral scanning. As before, $\eta_0$ denotes the maximum experimentally observed image quality enhancement under full spatial overlap and optimal alignment and $\chi$ is taken as 1.83.

%%%%% Bi
\bibliography{references.bib}